\documentstyle[12pt]{article}
\newcommand{\be}{\begin{equation}}
\newcommand{\ee}{\end{equation}}

\newcommand{\bea}{\begin{eqnarray}}
\newcommand{\eea}{\end{eqnarray}}
\newcommand{\leeq}{\lefteqn}
\newcommand{\non}{\nonumber}

\newcommand \ignore[1]{}
\newcommand{\PHNX}{\phi_{n}(\vec{x})}

\newcommand{\PHYX}{\phi_{Y}(\vec{x})}

\newcommand{\PSX}{\psi(\vec{x})}
\newcommand{\PSDX}{\psi^{\dagger}(\vec{x})}

\newcommand{\PPSX}{\Psi(X)}

\newcommand{\PPSDXD}{\Psi^{\dagger}(X')}

\newcommand{\Cno}{\hat{C}_{n}}
\newcommand{\Cmo}{\hat{C}_{m}}
\newcommand{\Cmdo}{\hat{C}^{\dagger}_{m}}
\newcommand{\Cndo}{\hat{C}^{\dagger}_n}

\newcommand{\ROK}{\tilde{\rho}(\vec{k})}
\newcommand{\ROKD}{\tilde{\rho}(\vec{k'})}

\newcommand{\VEX}{\vec{x}}
\newcommand{\VEXD}{\vec{x'}}

\newcommand{\VEK}{\vec{k}}
\newcommand{\VEKD}{\vec{k'}}

\newcommand{\LT}{\tilde{L}}

\newcommand{\DE}{\delta}
\newcommand{\PA}{\partial}
\newcommand{\TH}{\theta}

\newcommand{\JTV}{\tilde{J}(v)}
\newcommand{\DPV}{\partial\varphi(v)}

\newcommand{\AL}{\alpha}
\newcommand{\VP}{\varphi}

\begin{document}
\begin{titlepage}
\begin{normalsize}
\flushright{Revised version \\ UT-671 \\ March 1994} \\
\end{normalsize}
\vfil
\begin{LARGE}
\begin{center}
{$W_{\infty}$ algebra in the integer quantum Hall effects}
\end{center}
\end{LARGE}
\vfil
\begin{center}
\begin{Large}
{Hiroo Azuma}
\vfil
\end{Large}
{Department of Physics, Faculty of Science \\
University of Tokyo \\
7-3-1 Hongo, Bunkyo-ku, Tokyo 113, Japan } \\
\vfil
Abstract
\end{center}
We investigate the $W_{\infty}$ algebra in the integer
quantum Hall effects. Defining the simplest vacuum,
the Dirac sea, we evaluate the central extension
for this algebra. A new algebra which contains the central
extension is called the $W_{1+\infty}$ algebra.
We show that this $W_{1+\infty}$ algebra is an origin of
the Kac-Moody algebra which determines the behavior of edge
states of the system. We discuss the relation between
the $W_{1+\infty}$ algebra and the incompressibility of
the integer quantum Hall system.
\end{titlepage}
\vfill
\newpage
\renewcommand{\theequation}{\thesection.\arabic{equation}}
\setcounter{equation}{0}
\section{Introduction}
The quantum Hall effect(QHE) was discovered
by von Klitzing, Dorda and Pepper in 1980\cite{KvKGDMP}.
They were studying the Hall conductance of
a two-dimensional electron gas
in an inversion layer of a silicon metal-oxide-\\
semiconductor field-effect
transistor(MOSFET).
They noticed that the Hall conductance was quantized
at a very low temperature
under a high magnetic field.
Plotting the Hall conductance as a function of
the external magnetic field,
they obtained the plateaus where the Hall conductance
was equal to an
integer multiple of $e^{2}/h$ and the longitudinal
conductance was
essentially equal to zero.
We call this phenomenon the integer quantum Hall effect(IQHE)
today.

When a charged particle moves on a plane in a perpendicular
uniform
magnetic field, discrete energy levels appear.
They are called the Landau levels.
There are degenerate states in each level.
In the QHE whose filling factor $\nu$ is equal to or less
than one,
taking the limit of the large external magnetic field,
electrons are
constrained in the lowest Landau level(LLL)
\cite{REPSMG}\cite{EF}.
Because the QHE is a low temperature phenomenon, all spins of
electrons are aligned and we need not consider a freedom of
spins.
In the IQHE, the LLL is completely filled and
liquid of electrons is incompressible without an interaction.

The incompressibility is an important feature of the
QHE\cite{REPSMG}\cite{RBL}.
In an incompressible droplet, a local fluctuation of
the density is forbidden.
There are no gapless collective excitations in a droplet.
A gapless excitation can occur only at a deformation of
the edge.
This excitation is called an edge state\cite{RBLBIH}.
To investigate an edge state, Wen and Stone and others
considered
edge-charge operators.
It has been indicated that the dynamics of the edge is
described with a
one-dimensional chiral fermion and edge-charge operators satisfy
the Kac-Moody algebra\cite{MS}\cite{ABMS}\cite{XGW}\cite{JMMS}.

Recently, it has been shown that the infinite dimensional Lie
algebra
appears in the second-quantized system constrained to the LLL
\cite{SIDKBS}\cite{JMMS}.
It is the $W_{\infty}$ algebra.
Although the wave functions in the LLL could not form
a complete set, a second-quantized fermion field is
constructed from them.
Under unitary transformations of this field,
a wave function remains in the LLL
and the total fermion number does not change.
The generators of these transformations form the $W_{\infty}$
algebra.
If we define a vacuum as the state of a localized electron
droplet,
the $W_{\infty}$ algebra acquires the central extension and
non-trivial current operators appear\cite{ACGVDCATGRZ}.
In this paper, we derive the central extension for this
algebra by defining the
simplest vacuum, the Dirac sea.
The new algebra that we obtain contains the $U(1)$
Kac-Moody algebra.
We show that this Kac-Moody algebra describes the edge states
in the IQHE.

What we show in this paper is the following.
The $W_{\infty}$ algebra discussed in this paper contains
infinite
operators whose conformal dimensions are
equal to or above one.
We arrange these operators in order.
This $W_{\infty}$ algebra contains the $c=0$ Virasoro algebra.
We find primary fields which are made from the electron field.
We notice that the current operators contained in this
$W_{\infty}$ algebra commute with each other.(Commonly,
the $W_{\infty}$
algebra is defined as a closed algebra which contains infinite
operators
whose conformal dimensions are equal to or above two.
The $W_{\infty}$ algebra does not contain the current algebra.
And so, strictly speaking, the algebra that we are considering
now is not the $W_{\infty}$ algebra.
It is the $W_{1+\infty}$ algebra which does not contain
the central extension.)
Because there are no central extension terms in the
current algebra,
we can hardly find a physical meaning in it.
We notice that we had better define an appropriate vacuum and
obtain
a non-trivial current algebra.
Defining the simplest vacuum, the Dirac sea, we derive an
exact form of
the $W_{1+\infty}$ algebra which has central extension terms.
Furthermore, we construct this $W_{1+\infty}$ algebra
from a one-dimensional free fermion field.
This $W_{1+\infty}$ algebra contains the $c=1$ Virasoro algebra
and the $U(1)$ Kac-Moody algebra.
We can expect to extract some physical meanings from them.
Taking the classical limit (the limit of the large external
magnetic
field $B$ ), we show that electrons become an incompressible
droplet.
(The similar consideration is given in \cite{ACCATGRZ2}.)
We also show that the origin of the Kac-Moody algebra,
which is constructed from
edge-charge operators and controls the edge states,
is this $W_{1+\infty}$ algebra.
Using the bosonization transformation, we construct
the primary fields
and edge-charge operators from a free boson field.

Here, we summarize the contents of sections in this paper.
In section 2, we review a single-particle kinematics in the LLL.
The degenerate wave functions in the LLL can be labelled by
an operator which commutes with the Hamiltonian.
The wave functions in the LLL can be regarded as
a one-dimensional system.
Then, we review unitary transformations of the second-quantized
field and the $W_{\infty}$ algebra.
Particularly we consider the meaning of the large $B$ limit.
Although we can always eliminate the appearance
of the factor $B$ by
scaling the coordinates, there are some meanings in this limit.
We understand that this limit is a sort of the classical limit.
In section 3, we discuss some properties of the
$W_{\infty}$ algebra.
This algebra contains the Virasoro algebra.
We find its primary fields.
In section 4, we describe the $W_{\infty}$ algebra
with a one-dimensional free
fermion.

In section 5, we discuss a localized electron liquid.
We regard this state as a vacuum.
Defining the simplest vacuum, the Dirac sea,
we derive the central extension and
obtain an exact form of the $W_{1+\infty}$ algebra.
Using the result obtained in section 4, we construct
this $W_{1+\infty}$ algebra from a one-dimensional free fermion.
In this vacuum, the filling factor $\nu$ is equal to one.
It means that we are considering the IQHE.
In the $W_{1+\infty}$ algebra, there is a non-trivial
current algebra.
We can get some physical information from them.
Taking the classical limit $(B\gg 1)$,
we can regard electrons as a localized droplet
whose density is uniform.
Electrons behave as an incompressible droplet.

In section 6, we review the edge-charge operators defined by
Wen and Stone and others.
These operators form the Kac-Moody algebra.
We show that the origin of this Kac-Moody algebra is
the $W_{1+\infty}$
algebra.
In section 7, we describe edge-charge operators with a bosonized
representation.
We construct primary fields found in section 3 from a free boson.
We also construct an operator which adds a charge on the edge.


\setcounter{equation}{0}
\section{The many-body system in the lowest Landau level and the
$W_{\infty}$ algebra}

At first, we review a single-particle quantum mechanics
on a plane.
When an electron moves on a plane with a perpendicular
uniform magnetic field,
energy levels are quantized discretely.
These energy levels are called the Landau levels.
In the LLL, there are infinite degenerate states.
We can label these states perfectly by an appropriate operator
which commutes with the Hamiltonian.
And so, the degenerate states in the LLL can be regarded as
a one-dimensional system.

Then we consider a many-fermion system in the LLL.
We review the recent work
that the $W_{\infty}$ algebra appears
in unitary transformations of the second-quantized field
\cite{SIDKBS}\cite{JMMS}.
These transformations preserve the states in the LLL
and keep the total fermion number invariant.
Generators of these transformations are made from a density
operator and
the commutation relation of them is
the $W_{\infty}$ algebra.
We discuss the large $B$ limit.
We understand that the large $B$ limit is a sort of
the classical limit.

The Hamiltonian of a single-particle quantum mechanics is given by
\be
H_{0}=\frac{1}{2m}(\vec{p}+e\vec{A})^2,\;\;
\mbox{where}\;
\vec{p}=-i\vec{\nabla},\;
\;
\vec{\nabla}\times\vec{A}=B.
\ee
The unit is given by $c=\hbar=1$.
$B(>0)$ is a perpendicular uniform external magnetic field
on the $xy$-plane.
The momentum operators $\pi_{i}$ are given by
\be
\pi_{i}=(p+eA)_{i},\;\;
\mbox{for}\;
i=x,y,\;\;
[\pi_{x},\pi_{y}]=-ieB.
\ee
Annihilation and creation operators are defined in the form,
\be
a\equiv \frac{1}{\sqrt{2eB}} (\pi_{x}-i\pi_{y}),\;
\;
a^{\dagger}\equiv \frac{1}{\sqrt{2eB}} (\pi_{x}+i\pi_{y}),\;\;
[a,a^{\dagger}]=1.\label{eqn:AAcom}
\ee

Writing the Hamiltonian with the annihilation and creation
operators,
we can treat this system as a harmonic oscillator,
\be
H_{0}=\omega(a^{\dagger}a+\frac{1}{2}),\;
\;
E_{0}=\frac{1}{2}\omega,\frac{3}{2}\omega,\cdots,
\ee
where $\omega=eB/m$.
There are degenerate states in each level.
In the limit of the large $B$, the energy gap increases
and the electron eventually occupies the lowest level.
{}From now on we consider only the LLL.
The condition that $\phi(\VEX)$ is in the LLL is given by
$a\phi(\VEX)=0$.

Now, we define the useful operators,
\be
\hat X\equiv x-\frac{\pi_{y}}{eB},\;
\;
\hat Y\equiv y+\frac{\pi_{x}}{eB},\;\;
[\hat X,\hat Y]=\frac{i}{eB}. \label{eqn:XYcom}
\ee
There are following relations,
$[H_{0},\hat X]=[H_{0},\hat Y]=[H_{0},\hat X^{2}+\hat Y^{2}]=0$.
We can label the degenerate states in the LLL by eigenstates of
$\hat X,\hat Y$ or $\hat X^{2}+\hat Y^{2}$.

For example, let's consider wave functions in the LLL that
diagonalize
$\hat X^{2}+\hat Y^{2}$.
In the gauge $\vec{A}=(-By/2,Bx/2)$,
we obtain an orthonormal basis,
\be
\PHNX\equiv \sqrt{\frac{eB}{2\pi 2^{n}n!}}
(\bar z\sqrt{eB})^{n}\exp(-\frac{eB}{4}\vert z\vert^{2}),
\label{eqn:X2Y2fun}
\ee
where $z=x+iy$ and $n=0,1,2,\cdots$.
Because we consider only the LLL, we can regard
$\hat X^{2}+\hat Y^{2}$ as an angular momentum operator.

Taking the gauge $\vec{A}=(-By,0)$ and diagonalizing $\hat Y$,
we obtain another orthonormal basis of wave functions
in the LLL.
The basis is given by
\be
\PHYX\equiv\sqrt[4]{\frac{eB}{\pi}}\frac{1}{\sqrt{L}}
\exp\{ieBYx-\frac{eB}{2}(y-Y)^{2}\},\label{eqn:phiyx}
\ee
where $0\leq x<L$.
This wave function is localized around $y=Y$.
A density of states is equal to $eB/2\pi$.

Let's consider the many-body problem.
We define the filling factor by
$\nu=2\pi\rho/eB$, where $\rho$ is a density of electrons.
If $\nu\leq 1$ and $B\gg 1$, all the electrons are constrained
in the LLL.

In the following, we take the gauge $\vec{A}=(-By/2,Bx/2)$
and use the basis
of $\{\PHNX\}$ defined in $(\ref{eqn:X2Y2fun})$ for a while.
We perform the second-quantization in the following way,
\be
\PSX=\sum_{n=0}^{\infty}\Cno\PHNX.
\ee
$\Cno$ and $\Cndo$ obey the anti-commutation relation,
\be
\{\Cno,\Cmdo\}=\delta_{n,m}. \label{eqn:Ccom}
\ee
($\{\PHNX\}$ is not a complete set.
And so, $\PSX$ and $\PSDX$ do not obey the usual
anti-commutation relation.)

We consider a unitary transformation of $\{\Cno\}$,
\be
\Cno\rightarrow\Cno'=\sum_{m=0}^{\infty}u_{nm}\Cmo.
\ee
This transformation preserves the LLL condition,
$a\psi'(\VEX)=0$,
and keeps the total fermion number invariant,
$\int d^{2}\!x\,\DE\rho(\VEX)=0$, where
$\rho(\VEX)=\PSDX\PSX$.
The generator is given by the functional,
\be
\rho[\xi]\equiv\int d^{2}\!x\,\rho(\VEX)\xi(\VEX),\;\;
[-i\rho[\xi],\PSX]=\DE\PSX.
\ee

Let's consider the commutation relation of $\rho[\xi]$.
Because it is difficult to calculate the commutation
relation of $\rho[\xi]$
directly, we derive the commutation relation
of the Fourier component
of $\rho(\VEX)$ first.
Fourier components of the density operator $\rho(\VEX)$ are
given by
\bea
\ROK&=&\int d^{2}\!x\, \rho(\VEX)\exp(i\VEK\cdot\VEX) \non \\
&=&\sum_{n,m=0}^{\infty}\Cmdo\Cno
\frac{(-i\sqrt{2eB})^{n+m}}{\sqrt{n!m!}}
(\frac{\PA}{\PA\bar{k}})^{n}(\frac{\PA}{\PA k})^{m}
\exp(-\frac{|\vec{k}|^{2}}{2eB}). \label{eqn:ROKdef}
\eea
Using this representation, we obtain
\be
[\ROK,\ROKD]=-2i\,\sin(\frac{\VEK\times\VEKD}{2eB})
\tilde{\rho}(\VEK+\VEKD)\exp(\frac{\VEK\cdot\VEKD}{2eB}).
\label{eqn:ROKROKcom}
\ee

Here, we define $W(\VEK)$ in the form,
$W(\VEK)\equiv\ROK \exp(|\vec{k}|^{2}/4eB)$.
The commutation relation of $W(\VEK)$ is given by
\be
[W(\VEK),W(\VEKD)]=-2i\,\sin (\frac{\VEK\times\VEKD}{2eB})
W(\VEK +\VEKD).
\label{eqn:WKcom}
\ee
This commutation relation is called the $W_{\infty}$ algebra
or the Fairlie-Fletcher-Zachos algebra\cite{SIDKBS}\cite{JMMS}
\cite{IB}\cite{IIK}.
(Strictly speaking, this is the $W_{1+\infty}$ algebra which has no
central extension terms.)

In the limit of $B\gg 1$, we obtain
\be
[W(\VEK),W(\VEKD)]=-\frac{i}{eB}(\VEK\times\VEKD) W(\VEK +\VEKD).
\ee
This algebra is called the $w_{\infty}$ algebra
or the area-preserving diffeomorphisms\cite{ES}.
We can regard $(\ref{eqn:WKcom})$ as a quantized
version of the area-preserving diffeomorphisms.
Here we must pay attention to the following.
We can always eliminate the appearance of the factor $eB$
by scaling
the coordinates,
$\vec{x'}=\sqrt{eB/2}\VEX$.
Therefore, strictly speaking, there is no meaning
in taking the large $B$ limit.
But, we can interpret this limit as the classical limit which
reduces
$(\ref{eqn:WKcom})$ to the $w_{\infty}$ algebra.
The large $B$ limit sometimes reveals important information
contained in the theory.
(The classical limit is discussed in \cite{ACCATGRZ2}, too.
In section 5, we will understand that the electron liquid shows
the incompressibility under this limitation.
The same result is obtained in \cite{ACCATGRZ2}.)

A commutation relation of $\rho[\xi]$ is obtained,
\be
[\rho[\xi_{1}],\rho[\xi_{2}]]
=\rho[\{\{\xi_{1},\xi_{2}\}\}],\label{eqn:RXIRXIcom}
\ee
where
\be
\{\{\xi_{1},\xi_{2}\}\}
=-\sum_{n=0}^{\infty}\frac{(-1)^{n}}{n!}(\frac{2}{eB})^{n}
\{\partial^{n}\xi_{1}(\VEX)\bar{\partial}^{n}\xi_{2}(\VEX)
-\bar{\partial}^{n}\xi_{1}(\VEX)\partial^{n}\xi_{2}(\VEX)\}.
\ee
$\{\{\xi_{1},\xi_{2}\}\}$ is called the Moyal bracket\cite{IABS}.
In the limit of $B\gg 1$,
the Moyal bracket becomes the Poisson bracket,
\be
\{\{\xi_{1},\xi_{2}\}\}\stackrel{B\gg 1}{\rightarrow}
\frac{i}{eB}[\partial_{x}\xi_{1}\partial_{y}\xi_{2}
-\partial_{y}\xi_{1}\partial_{x}\xi_{2}]\equiv\{\xi_{1},
\xi_{2}\}_{PB}.
\ee

Even if we take the gauge $\vec{A}=(-By,0)$ and use the
$\{\phi_{Y}(\VEX)\}$ defined in $(\ref{eqn:phiyx})$,
we obtain the same commutation relations.

\setcounter{equation}{0}
\section{Some properties of the $W_{\infty}$ algebra}
Let's investigate some properties of the $W_{\infty}$ algebra
obtained in
$(\ref{eqn:WKcom})$.
This algebra contains infinite operators.
Labelling these operators with the
conformal dimensions, we arrange them in order.
We show that $(\ref{eqn:WKcom})$ contains the Virasoro algebra.
Then, we look for primary fields.

We define operators,
\be
\LT^{(n)}(k_{x})\equiv \exp(\frac{k_{x}^{2}}{4eB})
\int d^{2}\!x\, y^{n}
\exp(ik_{x}x)\rho(\VEX). \label{eqn:LTdef}
\ee
{}From $(\ref{eqn:WKcom})$, we obtain commutation
relations of $\LT^{(n)}(k_{x})$,
\bea
\leeq{
[\LT^{(n)}(k_{x}),\LT^{(m)}(k_{x}')] } \non \\
&=& n!m!(-i)^{n+m}(-2i)
\sum^{\infty}_{p,q,r=0}\sum^{2p+1}_{j=0}\sum^{r}_{h=0}
\delta_{n,2p+1-j+q+h}\delta_{m,j+q+r-h} \non \\
&&\times (-1)^{p+1-j}(i)^{r}
\frac{1}{(2eB)^{2p+1+q}}\frac{1}{j!(2p+1-j)!q!(r-h)!h!} \non \\
&&\times k_{x}^{j}k_{x}'^{2p+1-j}\LT^{(r)}(k_{x}+k_{x}').
\label{eqn:LTcom}
\eea

Because these commutation relations are complicated,
we discuss characteristic ones.
We redefine $\LT^{(n)}(k_{x})$ by the form,
\be
L^{(n)}_{m}\equiv(\frac{eBL}{2\pi})^{n}\LT^{(n)}(-\frac{2\pi m}{L}),
\;\;m\in\mbox{\boldmath$Z$}.
\ee
Some commutation relations can be written in the forms,
\be
[L^{(0)}_{n},L^{(0)}_{m}]=0,\label{eqn:KA}
\ee
\[
[L^{(0)}_{n},L^{(1)}_{m}]=nL^{(0)}_{n+m},
\]
\be
[L^{(1)}_{n},L^{(1)}_{m}]=(n-m)L^{(1)}_{n+m}, \label{eqn:VIRA}
\ee
\[
[L^{(0)}_{n},L^{(2)}_{m}]=2nL^{(1)}_{n+m},
\]
\[
[L^{(1)}_{n},L^{(2)}_{m}]=(2n-m)L^{(2)}_{n+m}
-n\{\frac{nm}{4}+(\frac{L}{2\pi})^{2}eB\}L^{(0)}_{n+m},
\]
\[
[L^{(2)}_{n},L^{(2)}_{m}]=2(n-m)L^{(3)}_{n+m}
+(n-m)\{\frac{nm}{2}+(\frac{L}{2\pi})^{2}2eB\}L^{(1)}_{n+m}.
\]
In these relations, $L^{(0)}_{n}$ looks like a lowering operator.
It is important that $L^{(0)}_{n}$ and $L^{(1)}_{n}$ form
a closed algebra.
$(\ref{eqn:VIRA})$ is the Virasoro algebra.

The conformal dimension is decided in the following way.
If an operator $A_{m}$ satisfy
\be
[L^{(1)}_{n},A_{m}]=\{n(h-1)-m\}A_{n+m}+\cdots,
\ee
$A_{m}$ is the Fourier mode of the field whose conformal
dimension is $h$.
We notice that $L^{(n)}_{m}$ is a component
of the conformal dimension $(n+1)$ field.

The algebra defined in $(\ref{eqn:WKcom})$ contains the
current operator $L^{(0)}_{n}$.
Commonly, the $W_{\infty}$ algebra contains infinite operators
whose
conformal dimensions are equal to or above two.
The $W_{\infty}$ algebra does not contain current operators.
Therefore, strictly speaking, the algebra defined in
$(\ref{eqn:WKcom})$
is not the $W_{\infty}$ algebra.
It is the the $W_{1+\infty}$ algebra which has not central
extension terms.
In particular, the algebra of the current operator $(\ref{eqn:KA})$
is trivial.
If we define a vacuum and consider a central extension term,
there must be an interesting physics in the current algebra.
In section 5, we define the simplest vacuum and derive the central
extension terms.

Next, we look for primary fields of $L^{(1)}_{n}$.
At first, we consider a field written in the form,
\be
\zeta(x)\equiv\int dy\,\psi(\VEX).
\ee
Let's take the gauge
$\vec{A}=(-By,0)$ and use the eigenfunctions of $\hat Y$
defined in $(\ref{eqn:phiyx})$.
We impose a periodic boundary condition, $0\leq X<L$.
$\zeta(x)$ is given in the form,
\be
\zeta(x)=\sum_{n}\Cno\sqrt{\frac{2}{L}}\sqrt[4]{\frac{\pi}{eB}}
\exp(ieBY_{n}x),\label{eqn:XIdef}
\ee
where $Y_{n}=2\pi n/eBL,\;n=0,\pm 1,\pm 2,\cdots.$
$\LT^{(1)}(k_{x})$ is given in the form,
\be
\LT^{(1)}(k_{x})=\frac{2\pi}{eBL}
\sum_{n}\Cndo\hat{C}_{n-Lk_{x}/2\pi}
\frac{1}{2}(2n-\frac{Lk_{x}}{2\pi}). \label{eqn:LTONEdef}
\ee
The commutation relations of $\LT^{(1)}(k_{x})$ and $\zeta(x)$,
$\zeta^{\dagger}(x)$ are
given by
\be
[\LT^{(1)}(k_{x}),\zeta(x)]=\frac{1}{eB}\exp(ik_{x}x)
(i\partial_{x}-\frac{k_{x}}{2})\zeta(x),
\ee
\be
[\LT^{(1)}(k_{x}),\zeta^{\dagger}(x)]=\frac{1}{eB}\exp(ik_{x}x)
(i\partial_{x}-\frac{k_{x}}{2})\zeta^{\dagger}(x).
\ee
Now, we define a new variable and new field operators,
\be
w\equiv \exp(-i\frac{2\pi}{L}x),\;\;
\eta(w)\equiv\frac{\zeta(x)}{\sqrt{w}},\;\;
\bar{\eta}(w)\equiv\frac{\zeta^{\dagger}(x)}{\sqrt{w}}.
\ee
We can express the commutation relations in the form,
\[
[L^{(1)}_{n},\eta(w)]=w^{n+1}\partial_{w}\eta(w)
+\frac{1}{2}(n+1)w^{n}\eta(w),
\]
\be
[L^{(1)}_{n},\bar{\eta}(w)]=w^{n+1}\partial_{w}\bar{\eta}(w)
+\frac{1}{2}(n+1)w^{n}\bar{\eta}(w).
\label{eqn:PRIMARY}
\ee
$\eta(w)$ and $\bar{\eta}(w)$ are primary fields.
Because we are considering free fermions, it is quite reasonable
that both of their conformal dimensions are equal to $1/2$.

\setcounter{equation}{0}
\section{The free fermion representation of the $W_{\infty}$
algebra}
In this section, we show that the $W_{\infty}$ algebra can be
constructed
from a\\
one-dimensional free fermion field\cite{EBCNPLJRESXS}.
In section 5, we will define the Dirac sea vacuum and
evaluate the central extension.
Then, we will show that the $W_{1+\infty}$ algebra, which contains
the central extension, is constructed from a one-dimensional
free fermion field, using the result obtained in this section.

At first, we obtain an explicit form of $L^{(l)}_{j}$.
Taking the gauge $\vec{A}=(-By,0)$ and using wave functions
defined in $(\ref{eqn:phiyx})$, we obtain general forms of
$\{L^{(l)}_{j}\}$,
\[
L^{(2l)}_{j}=\sum_{n}\Cndo\hat{C}_{n+j}\sum_{p=0}^{l}
\left(\begin{array}{c}2l\\2p\end{array}\right)
\frac{(2p-1)!!}{2^{2l+p}}(\frac{L}{\pi})^{2p}
(eB)^{p}(2n+j)^{2(l-p)},
\]
\be
\mbox{for}\;\;l=0,1,2,\cdots,
\ee
\[
L^{(2l+1)}_{j}=\sum_{n}\Cndo\hat{C}_{n+j}\sum_{p=0}^{l}
\left(\begin{array}{c}2l+1\\2p\end{array}\right)
\frac{(2p-1)!!}{2^{2l+p+1}}(\frac{L}{\pi})^{2p}(eB)^{p}
(2n+j)^{2(l-p)+1},
\]
\be
\mbox{for}\;\;l=0,1,2,\cdots.
\ee
Because these are complicated, we present some examples,
\be
L^{(0)}_{j}=\sum_{n}\Cndo\hat{C}_{n+j},
\;\;
L^{(1)}_{j}=\sum_{n}\Cndo\hat{C}_{n+j}\frac{1}{2}(2n+j).
\ee

We define operators,
\be
T^{(l)}(x)\equiv(\frac{1}{L})^{l+1}\sum_{j}L^{(l)}_{j}
\exp(i\frac{2\pi}{L}jx).
\ee
$T^{(1)}(x)$ is an energy-momentum tensor in the conformal
field theory(CFT).
We notice that $T^{(0)}(x)$ can be written in the form,
\be
T^{(0)}(x)=\frac{1}{2}\sqrt{\frac{eB}{\pi}}
\zeta^{\dagger}(x)\zeta(x),
\ee
where $\zeta(x)$ is the primary field defined in
$(\ref{eqn:XIdef})$.
Using $\zeta(x)$, we can rewrite $T^{(1)}(x)$
in the form,
\be
T^{(1)}(x)=\frac{1}{2}\sqrt{\frac{eB}{\pi}}\zeta^{\dagger}(x)
(-\frac{i}{4\pi})(\PA_{x}-\stackrel{\leftarrow}{\PA_{x}})\zeta(x),
\ee
The general form of $T^{(l)}(x)$ is given by
\bea
\leeq{
T^{(2l)}(x)=\frac{1}{2}\sqrt{\frac{eB}{\pi}}\zeta^{\dagger}(x)
\sum^{l}_{p=0}\left(\begin{array}{c}2l\\2p\end{array}\right)
\frac{(2p-1)!!}{2^{2l+p}}(\frac{\sqrt{eB}}{\pi})^{2p} }\non \\
&&\times\{-\frac{i}{2\pi}(\PA_{x}-\stackrel{\leftarrow}
{\PA_{x}})\}^{2(l-p)}
\zeta(x),\;\;\mbox{for}\;\;l=0,1,2,\cdots,
\eea
\bea
\leeq{
T^{(2l+1)}(x)=\frac{1}{2}\sqrt{\frac{eB}{\pi}}\zeta^{\dagger}(x)
\sum^{l}_{p=0}\left(\begin{array}{c}2l+1\\2p\end{array}\right)
\frac{(2p-1)!!}{2^{2l+p+1}}(\frac{\sqrt{eB}}{\pi})^{2p} }\non \\
&&\times\{-\frac{i}{2\pi}(\PA_{x}-\stackrel{\leftarrow}
{\PA_{x}})\}^{2(l-p)-1}
\zeta(x),\;\;\mbox{for}\;\;l=0,1,2,\cdots.
\eea

Conversely, if we prepare a one-dimensional free fermion field
$\zeta(x)$, we can always construct
$\{T^{(l)}(x)\}$ and $\{L^{(l)}_{j}\}$.
We can construct the $W_{\infty}$ algebra from $\zeta(x)$.

\setcounter{equation}{0}
\section{The central extension of the $W_{\infty}$ algebra for the
$\nu=1$ Dirac sea vacuum}
In this section, we consider the state of the localized electron
liquid.
We define a vacuum and derive a central
extension.
We obtain an exact form of the $W_{1+\infty}$ algebra.
The central extension creates the Kac-Moody current.
To make the problem easy, we assume the simplest vacuum,
the Dirac sea.
(The similar consideration is given in \cite{ACGVDCATGRZ}.)
Using the result obtained in section 4, we construct the
$W_{1+\infty}$ algebra
from a one-dimensional free fermion field.
We take the large $B$ limit of this vacuum.
We notice that the Dirac sea vacuum can be regarded as
a liquid of electrons
filled in the area $y\leq 0$.
The filling factor is equal to $1$ and the liquid of electrons
shows incompressibility.

It has been shown already that the system considered now can
be regarded as
a one-dimensional fermion system.
In this system, $\hat{X}$ plays the role of a coordinate
and $\hat{Y}$ plays the role of a momentum,
$[\hat{X},\hat{Y}]=i/eB$.
We use a set of eigenstates defined in $(\ref{eqn:phiyx})$,
$\hat{Y}\phi_{Y_{n}}(\VEX)=Y_{n}\phi_{Y_{n}}(\VEX)$.
The vacuum state $\vert G\rangle$ is defined,
\be
\vert G\rangle\equiv\prod_{n=0}^{\infty}\hat{C^{\dagger}_{-n}}
\vert 0\rangle
=\hat{C^{\dagger}_{0}}\hat{C^{\dagger}_{-1}}
\hat{C^{\dagger}_{-2}}\cdots
\vert 0\rangle,\label{eqn:DIRACVACdef}
\ee
where $\hat{C_{n}}\vert 0\rangle=0$,
for $\:\forall n\in \mbox{\boldmath$Z$}$.
In the vacuum state $\vert G\rangle$, all states with non-positive
momenta are filled. (See Fig.1.)
The normal ordering : : is defined by the following.
$\hat{C}_{n}(n>0)$ or $\hat{C}_{n}^{\dagger}(n\leq 0)$
are put on the right of $\hat{C}_{n}^{\dagger}(n>0)$ or
$\hat{C}_{n}(n\leq 0)$.
Whenever operators are exchanged, they are multiplied by $(-1)$.

Using the Wick's theorem, we can derive a new commutation relation,
\be
[:\!W(\VEK)\!:,:\!W(\VEKD)\!:]=
-2i\,\sin (\frac{\VEK\times\VEKD}{2eB})\{:\!W(\VEK +\VEKD)\!:
+\langle W(\VEK +\VEKD)\rangle\},
\label{eqn:WKNORMcom}
\ee
where $\langle\mbox{ }\rangle$ is a vacuum expectation.
The second term in the right hand side of $(\ref{eqn:WKNORMcom})$
is the central extension.

Let's derive the vacuum expectation.
We impose the periodic boundary condition, $0\leq X<L$.
$\langle W(\VEK)\rangle$ is given by the form,
\be
\langle W(\VEK)\rangle
=2\pi\delta_{k_{x},0}\frac{1}{L}\frac{1}{1-\exp(-i2\pi k_{y}/LeB)}.
\label{eqn:DIRACVAEX}
\ee
In the limit of $L\rightarrow\infty$, we obtain
\be
\langle W(\VEK)\rangle=\delta_{k_{x},0}\frac{eB}{ik_{y}}.
\ee
We obtain a new commutation relation\cite{SI},
\bea
\leeq{
[:\!W(\VEK)\!:,:\!W(\VEKD)\!:]
=-2i\,\sin(\frac{\VEK\times\VEKD}{2eB}):\!W(\VEK+\VEKD)\!:}\non \\
&&-\delta_{k_{x}+k'_{x},0}\frac{2eB}{k_{y}+k'_{y}}
\sin(\frac{\VEK\times\VEKD}{2eB}). \label{eqn:W1Kcom}
\eea

To understand $(\ref{eqn:W1Kcom})$ more clearly, we derive
commutation relations of\\
$:\!\LT^{(n)}(k_{x})\!:$,
\bea
\leeq{
[:\!\LT^{(n)}(k_{x})\!:,:\!\LT^{(m)}(k_{x}')\!:] } \non \\
&=&:[\LT^{(n)}(k_{x}),\LT^{(m)}(k_{x}')]:  \non \\
&&-\delta_{k_{x}+k'_{x},0}n!m!(-i)^{n+m}\sum^{\infty}_{r,q=0}
\sum^{r}_{j=0}\sum^{2q}_{h=0}(-1)^{r+q}2^{-r}(2eB)^{-2q-r}
k_{x}^{2q+1} \non \\
&&\times\frac{1}{j!(r-j)!h!(2q-h)!(2q+1)}\delta_{n,2j+h}
\delta_{m,2(r-j+q)-h}.
\eea
We obtain a new closed algebra of $:\!L^{(0)}_{n}\!:$ and
$:\!L^{(1)}_{n}\!:$,
\[
[:\!L^{(0)}_{n}\!:,:\!L^{(0)}_{m}\!:]=n\delta_{n+m,0},\;\;
[:\!L^{(0)}_{n}\!:,:\!L^{(1)}_{m}\!:]=n:\!L^{(0)}_{n+m}\!:,
\]
\be
[:\!L^{(1)}_{n}\!:,:\!L^{(1)}_{m}\!:]=(n-m):\!L^{(1)}_{n+m}\!:
+\delta_{n+m,0}\frac{1}{12}n^{3}. \label{eqn:vicom}
\ee
If we redefine $L^{(1)}_{0}$ in the form,
$L^{(1)}_{0}\rightarrow L^{(1)}_{0}-1/24$,
we can rewrite $(\ref{eqn:vicom})$ in the form,
\be
[:\!L^{(1)}_{n}\!:,:\!L^{(1)}_{m}\!:]=(n-m):\!L^{(1)}_{n+m}\!:
+\delta_{n+m,0}\frac{1}{12}n(n^{2}-1). \label{eqn:VIRAcom}
\ee
These are the $c=1$ Virasoro algebra and the $U(1)$ Kac-Moody
algebra.
The algebra defined in $(\ref{eqn:W1Kcom})$ is called the
$W_{1+\infty}$
algebra.
(The $W_{1+\infty}$ algebra is a closed algebra which contains
infinite operators whose conformal dimensions are equal to or
above one.)

In section 4, we showed that the $W_{\infty}$ algebra
can be constructed
from $\zeta(x)$.
It is true for the $W_{1+\infty}$ algebra, too.
We can construct $\{:\!L^{(l)}_{j}\!:\}$ in the similar way.
For example, we define $:\!T^{(0)}(x)\!:$ by the form,
\be
:\!T^{(0)}(x)\!:\equiv
\frac{1}{2}\sqrt{\frac{eB}{\pi}}:\!\zeta^{\dagger}(x)\zeta(x)\!:.
\label{eqn:KacMoodyCurrent}
\ee
We can construct the $W_{1+\infty}$ algebra from a one-dimensional
fermion field $\zeta(x)$.

Here, we consider the following.
We can construct the energy-momentum tensor from
the Kac-Moody current
$:\!T^{(0)}(x)\!:$ as the Sugawara form.
Decomposing this energy-momentum tensor, we obtain the Virasoro
operators.
This new Virasoro operator corresponds to $:\!L_{j}^{(1)}\!:$.

We can rewrite $(\ref{eqn:W1Kcom})$ by the functional,
\bea
\leeq{
[:\!\rho[\xi_{1}]\!:,:\!\rho[\xi_{2}]\!:] }\non \\
&=&:\rho[\{\{\xi_{1},\xi_{2}\}\}]:
+\frac{1}{2\pi}\int d^{2}\!x\int d^{2}\!x'\,
\delta(x-x')\delta(y)\delta(y')
\frac{i}{2}(\partial_{x}-\partial_{x'}) \non \\
&&\times\{1+\frac{1}{2eB}\frac{1}{2}(\vec{\nabla}^{2}+
\vec{\nabla}'^{2})+
(\frac{1}{2eB})^{2}[\frac{1}{8}(\vec{\nabla}^{2}+
\vec{\nabla}'^{2})^{2}
-\frac{1}{3!}(\vec{\nabla}\times\vec{\nabla'})^{2}]+\cdots\} \non \\
&&\times\xi_{1}(\VEX)\xi_{2}(\VEXD) \non \\
&=&:\rho[\{\{\xi_{1},\xi_{2}\}\}]:
+\frac{1}{2\pi}\int d^{2}\!x\int d^{2}\!x'\,
\delta(x-x')\delta(y)\delta(y')
\frac{i}{2}(\partial_{x}-\partial_{x'}) \non \\
&&\times\int du\,\delta(u-\frac{\vec{\nabla}
\times\vec{\nabla'}}{2eB})
\frac{\sin u}{u}\exp(\frac{\vec{\nabla}^{2}+\vec{\nabla'}^{2}}{4eB})
\xi_{1}(\VEX)\xi_{2}(\VEXD) \label{eqn:DIRACSEA}.
\eea
In the limit of $B\gg 1$, $(\ref{eqn:DIRACSEA})$
becomes simple.
We neglect $O(1/B)$ terms.
We obtain
\be
[:\!\rho[\xi_{1}]\!:,:\!\rho[\xi_{2}]\!:]
\simeq:\rho[\{\xi_{1},\xi_{2}\}_{PB}]:+
\frac{eB}{2\pi}\int d^{2}\!x\,
\TH(-y)\{\xi_{1},\xi_{2}\}_{PB}.\label{eqn:BLIMROXROXcom}
\ee

This result is interesting.
Because of $(\ref{eqn:WKNORMcom})$,
in the limit of $B\gg 1$, the commutation relation of
$:\!\rho[\xi]\!:$ is given by
\be
[:\!\rho[\xi_{1}]\!:,:\!\rho[\xi_{2}]\!:]
\stackrel{B\gg 1}{\rightarrow}
:\rho[\{\xi_{1},\xi_{2}\}_{PB}]:
+\int d^{2}\!x\langle\rho(\VEX)\rangle\{\xi_{1},\xi_{2}\}_{PB}.
\label{eqn:BLIMROXROXcom2}
\ee
Comparing $(\ref{eqn:BLIMROXROXcom})$ with
$(\ref{eqn:BLIMROXROXcom2})$, we notice that
$\langle \rho(\VEX)\rangle$ is equal to $eB\TH(-y)/2\pi$
in the limit of $B\gg 1$.
Let's see Fig.1.
If $B\gg 1$, $X$ and $Y$ are approximately equal to $x$ and $y$,
\[
\hat X=x-\frac{\pi_{y}}{eB}\simeq x,\;
\;
\hat Y=y+\frac{\pi_{x}}{eB}\simeq y.
\]
This result means the following.
In the limit of $B\gg 1$,
the electrons are localized in the area $y\leq 0$.
We can conclude that the Fermi liquid in this system is
approximately an incompressible fluid in the limit of $B\gg 1$.
In particular, $eB/2\pi$ is the density of the LLL.
In this system, the filling factor $\nu$ is equal to one.


\setcounter{equation}{0}
\section{The edge-charge operators and the Kac-Moody algebra}
In this section we review the edge-charge operator
which has been discussed by Stone, Wen and others
\cite{EF}\cite{MS}\cite{XGW}\cite{JMMS}.
These authors have claimed that the commutation relation of
the edge-charge
operators is the Kac-Moody algebra.
We show that the origin of this Kac-Moody algebra is
the $W_{1+\infty}$ algebra by using the result obtained
in section 5.
We will understand that the $U(1)$ Kac-Moody algebra
rules the behavior
of the edge state.

Let's consider electrons not only in an external magnetic
field but also
in a weak electrostatic potential $V(\VEX)$,
\[
V(\VEX)=\left\{
\begin {array}{lll}
E\Lambda & \Lambda<y \\
Ey & -\Lambda\leq y\leq\Lambda \\
-E\Lambda & y<-\Lambda
\end{array}
\right.
\]
where $1\gg\Lambda >0$,
$\Lambda=\mbox{const}$,
and
$0<E\ll 1\;,\;E=\mbox{const.}$
In this case, the electrostatic potential keeps the
electrons inside
the area, $y\leq 0$.
In the limit of $B\gg 1$, the ground state of this system becomes
approximately the Dirac sea vacuum.
Therefore we can use the results
which are obtained in section 5.

Using Stone's notation, an edge-charge operator is defined,
\be
j(x)\equiv\int dy\,g_{\Lambda}(y):\!\rho(\VEX)\!:,
\;\;\mbox{where}\;\;
g_{\Lambda}(y)=\left\{
\begin{array}{ll}
1 & -\Lambda\leq y\leq\Lambda \\
0 & y<-\Lambda\, ,\,\Lambda<y
\end{array}
\right..
\ee
We consider that $\Lambda$ is larger than the magnetic length,
$\Lambda\gg l_{0}$ where $l_{0}=1/\sqrt{eB}$.
$j(x)$ is the charge operator which lies on the edge $(x,0)$.
(See Fig.1.)

A functional is defined,
\be
j[f]\equiv\int dx\,f(x)j(x)=:\!\rho[f\cdot g_{\Lambda}]\!:,
\ee
where $f(x)$ is any non-singular function.
Using $(\ref{eqn:BLIMROXROXcom})$, in the limit of $B\gg 1$,
the commutation relation of $j[f]$ is given by
\bea
[j[f_{1}],j[f_{2}]]
&=&[:\!\rho[f_{1}g_{\Lambda}]\!:,
:\!\rho[f_{2}g_{\Lambda}]\!:] \non \\
&\simeq &
-\frac{i}{2eB}\int_{-\infty}^{+\infty}dx
\int_{-\Lambda}^{\Lambda}dy\,\frac{\PA}{\PA y}
:\!\rho(\VEX)\!:(f_{1}'f_{2}-f_{1}f_{2}') \non \\
&&+\frac{i}{4\pi}\int dx\,(f_{1}'f_{2}-f_{1}f_{2}').
\eea

Let's evaluate the first term in the right hand side of this
equation.
Using the basis defined in $(\ref{eqn:phiyx})$, the first term is
written explicitly in the form,
\bea
\leeq{
\frac{i}{2}\sqrt{\frac{eB}{\pi}}\frac{1}{L}
\sum_{n,m}:\!\Cndo\Cmo\!:
\int_{-\infty}^{+\infty}dx\,(f_{1}'f_{2}-f_{1}f_{2}')
\exp\{-ieB(Y_{n}-Y_{m})x\} }\non \\
&&\times\int_{-\Lambda}^{\Lambda}dy\,[2y-(Y_{n}+Y_{m})]
\exp\{-eB[y-\frac{1}{2}(Y_{n}+Y_{m})]^{2}
-\frac{eB}{4}(Y_{n}-Y_{m})^{2}\}.\non
\eea
Because of the Gaussian, we notice that dominant contributions
are given when
$2y\simeq Y_{n}+Y_{m}$.
On the other hand, because of the factor $[2y-(Y_{n}+Y_{m})]$,
these contributions vanish.
We can neglect this term.
We obtain
\be
[j[f_{1}],j[f_{2}]]
\simeq
\frac{i}{4\pi}\int dx\,[f_{1}'(x)f_{2}(x)-f_{1}(x)f_{2}'(x)].
\label{eqn:STONEEQ}
\ee
(Stone obtained this relation, too.
He didn't consider the central extension of $(\ref{eqn:W1Kcom})$.
He substituted the mean value of $\rho(\VEX)$ in the right
hand side
of $(\ref{eqn:RXIRXIcom})$.)

We can write the commutation relation formally in another form,
\be
[j[f_{1}],j[f_{2}]]
=\int dx\int dx'\,f_{1}(x)f_{2}(x')[j(x),j(x')].
\ee
Hence, we obtain the commutation relation of $j(x)$
in the limit of $B\gg 1$,
\be
[j(x),j(x')]=-\frac{i}{2\pi}\PA_{x}\DE(x-x'). \label{eqn:CURRENTcom}
\ee

There is the Kac-Moody algebra in $(\ref{eqn:CURRENTcom})$.
Let's decompose $j(x)$ in the Fourier components,
\be
\tilde{j}_{n}\equiv\int dx\,\exp(-i\frac{2\pi}{L}nx)j(x).
\ee
{}From $(\ref{eqn:CURRENTcom})$, we obtain the Kac-Moody algebra,
$[\tilde{j}_{n},\tilde{j}_{m}]=n\DE_{n+m,0}$.

We can obtain the following relation easily,
\be
\tilde{j}_{n}=\exp\{-\frac{1}{4eB}(\frac{2\pi n}{L})^{2}\}
:\!L^{(0)}_{n}\!:.
\ee
In the limit of $B\gg 1$, $\tilde{j}_{n}$ is identified with
$:\!L^{(0)}_{n}\!:$.
Furthermore, we pay attention to the following.
{}From the Kac-Moody algebra, we can always construct
the $c=1$ Virasoro algebra.
We will discuss it in section 7.

Let's consider the physical meaning of $\tilde{j}_{n}$.
For example, in the limit of $B\gg 1$, $\tilde{j}_{-n}$ $(n>0)$
is given by
$\tilde{j}_{-n}\simeq
\sum_{l=-\infty}^{+\infty}:\!\hat{C_{l}^{\dagger}}\hat{C_{l-n}}\!:$.
This operator acts on the droplet in the following way.(See Fig.2.)
$:\!\hat{C_{l}^{\dagger}}\hat{C_{l-n}}\!:$ $(n\geq l>0)$
annihilates the
electron at $Y=Y_{l-n}$ and creates the electron at $Y=Y_{l}$.
Therefore, $\tilde{j}_{-n}$ shifts electrons along the direction
of $Y$
by a distance $2\pi n/eBL$.

Because the droplet is incompressible, $\tilde{j}_{-n}$ is an
operator
which deforms the edge of the droplet.
This deformation propagates with the velocity $v_{F}=E/B$
along the edge.
There is the current which is not parallel to the external
electric field $E$.
This is the IQHE.
The electric current $\vec{j}$ is given by
\[
\vec{j}=
\left(
\begin{array}{cc}
0&e^{2}/2\pi\\
-e^{2}/2\pi&0\\
\end{array}
\right)
\left(
\begin{array}{ll}
0\\
E\\
\end{array}
\right).
\]
Stone, Wen and others derived the Kac-Moody algebra by quantizing
the classical theory.
They quantized the classical equation of the surface wave on the
droplet
and obtained the one-dimensional chiral free fermion.

Generally we can conjecture the following.
We assume an arbitrary vacuum. (See Fig.3.)
A droplet of electrons is localized in the area, $S$.
$\PA S$ is an edge of $S$.
Taking the classical limit, we obtain
\bea
[:\!\rho[\xi_{1}]\!:,:\!\rho[\xi_{2}]\!:]
&\simeq &\int d^{2}\!x\,\langle\rho(\VEX)
\rangle\{\xi_{1},\xi_{2}\}_{PB} \non \\
&\simeq &\frac{i}{2\pi}\int_{S}d^{2}\!x\,\vec{\nabla}
\times\vec{\xi} \non \\
&\simeq &\frac{i}{2\pi}\oint_{\PA S}d\vec{x}\cdot\vec{\xi},
\label{eqn:GEneLIMcom}
\eea
\be
\mbox{where}\;\;
\rho(\VEX)=
\left\{
\begin{array}{ll}
eB/2\pi & \VEX\in S \\
0 & \VEX\not\in S
\end{array}
\right.,\;\;
\vec{\xi}=(\xi_{2},\xi_{1}).
\ee
{}From this equation, we notice that this commutation relation
reflects a behavior of the edge.


\setcounter{equation}{0}
\section{The bosonization of the one-dimensional
fermion system}
The bosonization transformation is known to be an important
property of the
one-dimensional fermion system\cite{EF}.
In this section, we express the Virasoro algebra and the Kac-Moody
algebra
with a boson field\cite{ABMS}\cite{XGW}.
It is a basic exercises of the CFT\cite{PG}.
We construct the primary fields defined in section 3 from a boson
field.

We define a current operator,
\be
\Omega(X)\equiv\Psi^{\dagger}(X)\Psi(X),
\ee
where $\PPSX\equiv\sum_{n}\Cno\langle X\vert n\rangle$,
$\hat{Y}\vert n\rangle\equiv Y_{n}\vert n\rangle$,
and $\langle X\vert n\rangle=(1/\sqrt{L})\exp(i2\pi nX/L)$.
$\PPSX$ satisfies the usual anti-commutation relation,\\
$\{\PPSX,\PPSDXD\}=\delta(X-X')$.

Here, we take the Dirac sea vacuum discussed in section 5.
$\Omega(X)$ is decomposed into a singular part and a non-singular
part,
\[
\Omega(X)=J(X)+\lim_{\epsilon\rightarrow 0}
\langle\Psi^{\dagger}(X+\epsilon)\Psi(X-\epsilon)\rangle,
\]
\be
\mbox{where}\;J(X)=:\!\Psi^{\dagger}(X)\Psi(X)\!:.
\ee
By the similar derivation of $(\ref{eqn:DIRACVAEX})$, the second
term
which has a singularity in the limit of $\epsilon\rightarrow 0$ is
given by
\be
\langle\Psi^{\dagger}(X+\epsilon)\Psi(X-\epsilon)\rangle
=\frac{i}{4\pi\epsilon}.
\ee
The commutation relation of $J(X)$ is obtained
in the form,
\be
[J(X),J(X')]
=-\frac{i}{2\pi}\PA_{X}\DE(X-X').
\ee
We find the $U(1)$ Kac-Moody algebra.
If we define $\{J_{n}\}$ in the form,
\be
J_{n}\equiv\int dX\,\exp(-i\frac{2\pi}{L}nX)J(X)
=\sum_{m}:\!\Cmdo\hat{C}_{m+n}\!:,
\ee
they satisfy $[J_{n},J_{m}]=n\DE_{n+m,0}$ and
$J^{\dagger}_{n}=J_{-n}$.

Let's consider how to define the normal ordering of $\{J_{n}\}$.
For $n>0$, $J_{n}$ annihilates the Dirac sea vacuum
$\vert G\rangle$,
\bea
J_{n}\vert G\rangle
&=&(-1)^{P}\sum_{m\geq -n+1}\Cmdo\hat{C}^{\dagger}_{0}
\hat{C}^{\dagger}_{-1}\hat{C}^{\dagger}_{-2}\cdots
\hat{C}_{m+n}\vert 0\rangle \non \\
&&-(-1)^{P'}\sum_{m\leq -n}\hat{C}_{m+n}\hat{C}^{\dagger}_{0}
\hat{C}^{\dagger}_{-1}\cdots\Cmdo\Cmdo\cdots\vert 0\rangle \non \\
&=&0,
\eea
where $P$ and $P'$ are proper integers.
The normal ordering $:\mbox{ }:$ is defined in the following way,
$:\!J_{n}J_{-n}\!:=J_{-n}J_{n}$ for $n>0$.

Using $\{J_{n}\}$, we can construct the $c=1$ Virasoro algebra.
$J(X)$ is written in the form,
\be
J(X)=\frac{1}{L}\sum_{n}\exp(i\frac{2\pi}{L}nX)J_{n}.
\ee
With a new variable
$v\equiv \exp(-i2\pi X/L)$,
we define a vector current $\JTV$,
\be
\JTV\,dv\equiv -i2\pi J(X)\,dX.
\ee
$\JTV$ is written as $\JTV=\sum_{n}J_{n}v^{-n-1}$.

We define $\{L_{n}\}$ in the following way,
\be
\frac{1}{2}:\!\JTV\JTV\!:=\sum_{n}L_{n}v^{-n-2}.
\ee
We have the $c=1$ Virasoro generators $\{L_{n}\}$,
\be
L_{n}=\frac{1}{2}\sum_{m}:\!J_{m}J_{n-m}\!:,
\ee
which satisfy the algebras,
\[
[L_{n},L_{m}]=(n-m)L_{n+m}+\frac{1}{12}n(n^{2}-1)\DE_{n+m,0},\;\;
[L_{n},J_{m}]=-mJ_{n+m}.
\]

We define a Bose field $\varphi(v)$ by
\be
i\PA_{v}\VP(v)\equiv \JTV.
\ee
We notice that this system is described by the CFT
which has an energy momentum tensor,
$T(v)=-(1/2):\!\DPV\DPV\!:=\sum_{n}L_{n}v^{-n-2}$.
$\varphi(v)$ is expanded in the form,
\be
\varphi(v)=-i\{\sum_{n\neq 0}-\frac{1}{n}J_{n}v^{-n}
+J_{0}\log v+K_{0}\},
\ee
where
$[J_{0},K_{0}]=1$, $K_{0}^{\dagger}=-K_{0}$.
$\DPV$ is a primary field of a conformal dimension $h=1$.

We define an exponential field $V_{\alpha}(v)$,
\be
V_{\AL}(v)\equiv:\!e^{i\AL\VP(v)}\!:.
\ee
$V_{\AL}(v)$ is a primary field
of a conformal dimension $h=\AL^{2}/2$.
Is $V_{\AL}(v)$ a boson or a fermion?
We give $V_{\AL}(v)$ a transparent form,
\bea
V_{\AL}(v)&=&
\exp(-\frac{1}{2}\AL^{2}\log v-\frac{\AL^{2}}{2n})
\prod_{n=1}^{\infty}\exp(\frac{\AL}{n}J_{-n}v^{n})
\exp(\AL J_{0}\log v)\exp(\AL K_{0}) \non \\
&&\times\prod_{m=1}^{\infty}\exp(-\frac{\AL}{m}J_{m}v^{-m}).
\eea
After some calculation, we get
\be
V_{\AL}(v)V_{\AL}(w)=(-1)^{\AL^{2}}V_{\AL}(w)V_{\AL}(v).
\ee
$\AL^{2}$ decides whether $V_{\AL}(v)$ is a boson or a fermion.
If we take $\AL=\pm 1$, we get a primary field which is
a fermion and has a conformal dimension $h=1/2$.
$V_{1}(v)$ and $V_{-1}(v)$ have appeared as $\eta(w)$
and $\bar{\eta}(w)$ in section 3 already.

We have the relations,
\be
[J(X),V_{\AL}(\exp(-i\frac{2\pi}{L}X'))]
=\AL\DE(X-X') V_{\AL}(\exp(-i\frac{2\pi}{L}X')),
\ee
\be
[J(X),V_{\AL}^{\dagger}(\exp(-i\frac{2\pi}{L}X'))]
=-\AL\DE(X-X') V_{\AL}^{\dagger}(\exp(-i\frac{2\pi}{L}X')).
\ee
We can interpret these equations as the following.
If we assume that $J(X)$ is a charge density operator at $X$ and
$\vert J(X)\rangle$ is its eigenstate,\\
$V_{\AL}(\exp(-i2\pi X'/L))$ adds a charge $\AL$ at $X'$,
\be
V_{\AL}(\exp(-i\frac{2\pi}{L}X'))\vert J(X)\rangle
\propto \vert J(X)+\AL\DE(X-X')\rangle,
\ee
and $V^{\dagger}_{\AL}(\exp(-i2\pi X'/L))$ adds a charge
$-\AL$ at $X'$,
\be
V^{\dagger}_{\AL}(\exp(-i\frac{2\pi}{L}X'))\vert J(X)\rangle
\propto \vert J(X)-\AL\DE(X-X')\rangle.
\ee

An operator which adds the charge
$(1/2\pi)\PA\TH(X)/\PA X$ at X is given by
\[
:\!\exp\{\frac{i}{2\pi}\int dX'\,
\frac{\PA\TH(X')}{\PA X'}\VP(\exp(-i\frac{2\pi}{L}X'))\}\!:
=:\!\exp\{i\int dX'\,\TH(X')J(X')\}\!:.
\]
This operator has been constructed by Stone\cite{MS}.

\setcounter{equation}{0}
\section{Discussion and Acknowledgements}

It is indicated that there is a close relation between the QHE and
a one-dimensional system.
In this paper, we have presented a simple example of that relation.
We considered the $W_{\infty}$ algebra
in $\nu=1$ case.
The $W_{\infty}$ algebra in $\nu=1/m$ case has been discussed by
Karabali\cite{DK}.
The general extension of the $W_{\infty}$ algebra to
the $W_{1+\infty}$ algebra has been done recently by Kac and Radul
\cite{VKAR}.
Its application to the QHE is discussed by Cappelli, Trugenberger,
and Zemba\cite{ACCATGRZ}.

I am grateful to Prof. Eguchi for his guidance during my
graduate course.
I am also grateful to Dr. Iso for many valuable
discussions and suggestions.



\begin{thebibliography}{99}
\bibitem{KvKGDMP}K.v.Klitzing, G.Dorda and M.Pepper,
Phys.Rev.Let. {\bf 45},
494 (1980).
%
\bibitem{REPSMG}R.E.Prange and S.M.Girvin,
{\it ``The Quantum Hall Effects''}, Springer, New York, 1990.
%
\bibitem{EF}E.Fradkin,
{\it ``Field theories of condensed matter systems''},
Addison-Wesley, 1991.
%
\bibitem{PG}P.Ginsparg,
{\it ``Fields, Strings, and Critical Phenomena''},
Les Houches, Session XLIX, 1988, course 1.
{\it ``Applied conformal field theory''},
Elsevier Science, 1989.
%
\bibitem{RBL}R.B.Laughlin, Phys.Rev.B {\bf 27}, 3383 (1983);\\
R.B.Laughlin, Phys.Rev.Lett. {\bf 50}, 1395 (1983).
%
\bibitem{RBLBIH}R.B.Laughlin, Phys.Rev.B {\bf 23}, 5632 (1981);\\
B.I.Halperin, Phys.Rev.B {\bf 25}, 2185 (1982).
%
\bibitem{MS}M.Stone, Ann.Phys.(N.Y.) {\bf 207}, 38 (1991).
%
\bibitem{ABMS}A.Balatsky and M.Stone, Phys.Rev.B {\bf 43},
8038 (1991).
%
\bibitem{XGW}
X.G.Wen, Phys.Rev.B {\bf 41}, 12838 (1990); \\
X.G.Wen, Int.J.Mod.Phys.B {\bf 6}, 1711 (1992).
%
\bibitem{SIDKBS}
S.Iso, D.Karabali and B.Sakita, Phys.Lett.B {\bf 296}, 143 (1992);\\
A.Cappelli, C.A.Trugenberger and G.R.Zemba, Nucl.Phys.B {\bf 396},
465 (1993).
%
\bibitem{JMMS}J.Mart\'{i}nez and M.Stone,
{\it ``Current operators in the lowest Landau level''},
NSF-ITP-93-38, ITP Santa Barbara preprint.
%
\bibitem{IB}
D.B.Fairlie, P.Fletcher and C.K.Zachos, Phys.Lett.B {\bf 218},
203 (1989);\\
J.Hoppe, Phys.Lett.B {\bf 215}, 706 (1988); \\
I.Bakas, Phys.Lett.B {\bf 228}, 57 (1989); \\
C.N.Pope, L.J.Romans and X.Shen, Phys.Lett.B {\bf 236}, 173 (1990).
%
\bibitem{IIK}
The $W_{\infty}$ symmetry in the $2+1$ dimensional gauge theory
with the Chern-Simons term is discussed in\\
I.I.Kogan Mod.Phys.Lett.A {\bf 7}, 3717 (1992).
%
\bibitem{IABS}
D.B.Fairlie and C.K.Zachos, Phys.Lett.B {\bf 224}, 101 (1989);\\
P.Fletcher, Phys.Lett.B {\bf 248}, 323 (1990); \\
I.A.B.Strachan, Phys.Lett.B {\bf 283}, 63 (1992).
%
\bibitem{EBCNPLJRESXS}E.Bergshoeff, C.N.Pope, L.J.Romans, E.Sezgin
and X.Shen, \\
Phys.Lett.B {\bf 245}, 447 (1990).
%
\bibitem{ES}E.Sezgin,
{\it ``Area-Preserving Diffeomorphisms, $w_{\infty}$ Algebras and
$w_{\infty}$ Gravity''},
CTP-TAMU-13/92, preprint.
%
\bibitem{ACCATGRZ2}
A.Cappelli, C.A.Trugenberger and G.R.Zemba, Phys.Lett.B {\bf 306},
100 (1993).
%
\bibitem{SI}S.Iso, unpublished (1992).
%
\bibitem{ACGVDCATGRZ}
A.Cappelli, G.V.Dunne, C.A.Trugenberger and G.R.Zemba,
Nucl.Phys.B {\bf 398}, 531 (1993).
%
\bibitem{DK}D.Karabali,
{\it ``Algebraic Aspects of the Fractional Quantum Hall Effect''},
SU-4240-553, preprint.
%
\bibitem{VKAR}V.Kac and A.Radul,
{\it ``Quasifinite highest weight modules over the Lie algebra
of differential
operators on the circle''}, preprint.
%
\bibitem{ACCATGRZ}A.Cappelli, C.A.Trugenberger and G.R.Zemba,
{\it ``Classification of Quantum Hall University Class by
$W_{1+\infty}$
symmetry''}, MPI-Ph/93-75, DFTT 65/93, preprint.
%
\end{thebibliography}
\end{document}